\documentclass[onecolumn,nofootinbib]{revtex4}
\usepackage{setspace}
\usepackage{amsmath,amssymb}
\usepackage{graphicx}
\usepackage{dcolumn}
\usepackage{bm,color}
\usepackage{hyperref}
\usepackage{accents}
\usepackage{amssymb,float}
\usepackage{amsmath}
\usepackage{multirow}
\usepackage{tikz}
\usepackage{comment}
\usepackage{subcaption}
\usepackage[rightcaption]{sidecap}
\sidecaptionvpos{figure}{t}
\usepackage{enumerate}

\newcolumntype{P}[1]{>{\centering\arraybackslash}p{#1}}
\newcolumntype{M}[1]{>{\centering\arraybackslash}m{#1}}
\hypersetup{
    colorlinks=true,
    linkcolor=red,
    filecolor=magenta,      
    citecolor=blue
}

\newcommand{\dd}{\mathrm{d}}

\begin{document}

\title{Renormalization Group-Improved Gravitational Action: A Lagrangian Framework}

\author{Alfio Bonanno}
\email{alfio.bonanno@inaf.it}
\affiliation{INAF, Osservatorio Astrofisico di Catania, via S.Sofia 78, I-95123 Catania, Italy}

\author{Konstantinos F. Dialektopoulos}
\email{kdialekt@gmail.com}
\affiliation{Department of Mathematics and Computer Science, Transilvania University of Brasov, 500091, Brasov, Romania}

\author{Vasilios Zarikas}
\email{vzarikas@uth.gr}
\affiliation{Department of Mathematics, University of Thessaly, 35100, Lamia, Greece}

\begin{abstract}

A new approach for embedding the renormalization group running of Newton's constant and cosmological constant in gravity is proposed. This approach is based on a gravitational Lagrangian that gives rise to a new class of modified gravity theories where \( G \) and \( \Lambda \) are spacetime-dependent functions. The Lagrangian formulation can be interpreted as an effective gravitational action that encapsulates the scale dependence of \( G \) and \( \Lambda \), arising  from quantum effects in the early universe. We show that the new formalism can be discussed using partially the framework and results of Horndeski modified gravity, excluding the equations of motion of the scalar field. The study explores aspects of this new gravity action. We also analyze an interesting non-singular cosmological solution featuring power-law inflation and we discuss the generation of scalar and tensor perturbations within this framework.
\end{abstract}

\maketitle

\section{Introduction}\label{sec:intro}

Cosmology is a key area for studying quantum gravity, which could answer fundamental questions such as the characteristics of the Universe's expansion, the avoidance of initial singularities, and the Cosmological Constant problem. In recent years the asymptotic safety scenario has emerged as a promising framework for these issues. In this approach, the ultraviolet behaviour of quantum gravity is controlled by a fixed point at a non-zero value of the dimensionless coupling constant, resulting in an antiscreened, weaker Newton's constant at higher energies. Non-perturbative renormalization group (RG) equations predict that the dimensionless cosmological constant reaches a non-gaussian fixed point (NGFP) in the infinite cutoff limit, making the full Einstein-Hilbert Lagrangian renormalizable at a non-perturbative level \cite{Weinberg:1980gg}.

The introduction of the effective average action \cite{Berges:2000ew} and its functional RG equation for gravity \cite{Reuter:1996cp} has enabled detailed investigations of Newton's constant scaling behavior. This approach defines a Wilsonian RG flow on a theory space, comprising all diffeomorphism-invariant functionals of the metric. The emerging theory is not simply a quantization of classical general relativity; instead, its bare action corresponds to a nontrivial fixed point of the RG flow, making it a prediction rather than a classical quantization. The effective average action offers crucial advantages over other continuum implementations of the Wilson RG. Specifically, it is closely related to the standard effective action and defines a family of effective field theories $\Gamma_k [g_{\mu\nu} ]$ labeled by the coarse graining scale $k$. This property allows for a more direct extraction of physical information from the RG flow, especially in single-scale cases: if a physical process or phenomenon involves only a single typical momentum scale $p_0$, it can be described by a tree-level evaluation of $\Gamma_k[g_{\mu\nu}]$  with $k=p_0$. The precision of this effective field theory description depends on the size of the fluctuations relative to the mean values. If the fluctuations are large or if multiple scales are involved, it may be necessary to go beyond the tree-level analysis. 

The qualitative scale dependence of Newton's constant can be understood through the following physical reasoning. Consider that, at large distances, the primary quantum effects on geometry are described by quantizing the linear fluctuations of the metric. This results in a minimally coupled theory within a curved background spacetime, where the elementary particles, gravitons, possess energy and momentum. The vacuum of this theory is filled with virtual graviton pairs, and the challenge is to comprehend how these virtual gravitons react to a perturbation caused by an external test body placed in the vacuum. Assuming gravity remains universally attractive in this scenario, the gravitons will be drawn toward the test body. Consequently, the test body will be ``dressed'' by a cloud of virtual gravitons, making its effective mass, as observed from a distance, greater than it would be without any quantum effects \cite{Polyakov:1993tp}.

In quantum gravity, unlike in QED where quantum fluctuations screen external charges, these fluctuations have an antiscreening effect on external test masses. This implies that Newton’s constant, $G(k)$ becomes a scale-dependent quantity which tends to zero at small distances as $k\sim 1/r$. In fact, in recent years, substantial evidence has accumulated supporting the antiscreening behavior of Newton's constant at high energies \cite{Niedermaier:2006wt,Codello:2008vh,Percacci:2017fkn,Eichhorn:2018yfc,Bonanno:2020bil, Reichert:2020mja}. Clearly, having complete control over the renormalized flow up to 
$k=0$ would entail understanding the quantum effective action. Unfortunately, this task remains significantly distant from achievement. Thus, the best approach available is to employ ``renormalization group improvement'' to derive a qualitative approximation to the full quantum effective action. This method enables predictions within the strong field regime and has been extensively applied in cosmology \cite{Bonanno:2017pkg,Good:2023xwp, Kofinas:2016lcz} and black hole physics \cite{Platania:2023srt, Kofinas:2015sna}.

The screening behavior in QED is well-understood, but it is useful to recall how this result is obtained using ``renormalization group improvement'', a common technique in particle physics to incorporate dominant quantum corrections into the Born approximation of a scattering cross-section. Starting from the classical potential we replace the charge with the running charge $e^2\rightarrow e^2(k)$ with $k\sim 1/r$ and we obtained the correct (one-loop, massless) Uehling potential \cite{PhysRev.48.55}, typically derived through conventional perturbative methods. 

The challenge arises when attempting to apply the same methodology in gravity, particularly when expressing $k$ as $k\sim 1/\ell (x^\mu)$, where $\ell(x^\mu)$ represents a characteristic length for the propagation of fluctuations with energy $k$. This difficulty stems from the fact that the flow equation, by its very construction, maintains diffeomorphism invariance at any $k$ thus remaining agnostic to the background field metric used to project onto a finite-dimensional subset of the ``theory space."

Various approaches exist to address this challenge. The first involves selecting a fiducial metric, typically a solution of the Einstein equations, and enhancing it through RG improvement by substituting the Newton constant $G$ with its running counterpart, while also implementing a cutoff identification akin to $k\sim 1/r$. However, this method's limitation lies in the possibility that the enhanced metric may not precisely conform to the Einstein equations. Yet, one can conceptualize this approach as a form of ``Thomas-Fermi'' approximation, focusing primarily on leading quantum corrections. The improved $g_{\mu\nu}$ metric then serves as an ``emergent'' description of spacetime geometry, contingent upon the scale-dependent behavior of the Newton constant.

Another approach is to focus on the energy scale $k$ associated directly with the field strength rather than an observational scale. This approach is inspired by the analogy with the cases of QED and QCD, where higher-order contributions to the Uehling potential are acquired through the renormalization group improvement of the QED action. In this method, the field strength $F_{\mu\nu} F^{\mu\nu}$ is used as a cutoff instead of $1/r$. The drawback with this approach in gravity that Kretschmann scalar is not in general positive definite and it is difficult to associate an invariant meaning of distance to this object. On the other hand in some cases it has proven to be useful to identify the relevant energy scale with the Ricci scalar $R$ via $k^2 \propto R$ \cite{Bonanno:2012jy}.

In recent years, various endeavors have been undertaken to introduce a scale-dependent formulation of gravitational action that upholds diffeomorphism invariance at every scale $k$. These efforts encompass using either a Brans-Dicke Lagrangian approach  \cite{Reuter:2003ca} or a Hamiltonian formulation \cite{Bonanno:2004ki}. However, the resulting scale-dependent equations impose stringent constraints on potential scale identifications as both $G$ and $\Lambda$ carry energy and momentum in general. A recent study \cite{Bonanno:2020qfu} presented the most comprehensive class of field equations with variable $G$ and $\Lambda$, which are consistent with diffeomorphism invariance and contain up to second order derivatives of the metric, $G$, and $\Lambda$. Solutions were derived for various plausible RG trajectories. In this work, we aim to extend the findings of \cite{Bonanno:2020qfu} and propose a potential Lagrangian description of variable $G$ and $\Lambda$ gravity. Particularly, we will demonstrate that non-singular cosmologies with power-law inflation naturally emerge in this context due to the running near the NGFP.

The paper is organized as follows. In Sec.~\ref{sec:modified ein-eq} we review the results found in \cite{Bonanno:2020qfu}, i.e. the effective field equations for a covariant theory with varying Newton's and cosmological constant general functions of spacetime. In Sec.~\ref{sec:constructing-action} we construct a suitable action which generates the above effective equations of motion. The rest of the paper explores the properties of this action. In Sec.\ref{App}, we present a quantum cosmology analysis as well as non singular cosmological solutions following one specific RG flow. Finally, in Sec.\ref{PPS} we perform scalar and tensor perturbations working generally with our action and consequently applying them to one non singular cosmological solution. The metric signature is mostly plus and $c = 1 = \hbar$ is considered throughout the paper.

\section{Modified Einstein Equations}
\label{sec:modified ein-eq}

Authors in \cite{Bonanno:2020qfu}, found a new set of field equations with varying $G$ and $\Lambda$, that are mathematically and physically consistent. They are the most general equations containing up to second order derivatives in the metric, $G$ and $\Lambda$, under some mild assumptions.

Motivated by the concept of RG approaches to Quantum Gravity like Asymptotic Safety (AS), a spacetime-dependent cosmological, $\Lambda(x)$, and Newton's, $G(x)$, constants have to be introduced to the effective low energy Einstein equations. For presentation simplicity $x$ denotes $x^{\mu}$. Thus, we have $G_{\mu\nu}=-\Lambda(x) g_{\mu\nu}+8\pi G(x) T_{\mu\nu}$.  It becomes immediately apparent that more terms have to be included to satisfy the Bianchi identities  $G_{\mu\nu}{^{;\mu}}=0$. In the absence of extra kinetic terms in the equations of motion Bianchi identities lead to $8\pi(GT_{\mu\nu}){^{;\mu}}=\Lambda_{;\nu}$ which is inconsistent in the absence of matter ($T_{\mu\nu}=0$). Thus, additional covariant derivatives of $G$, $\Lambda$, and perhaps $T_{\mu\nu}$ have to be included. 
Furthermore, an assumption has been made in \cite{Bonanno:2020qfu} regarding $G$ and $\Lambda$ that are not allowed to change signs during their evolution (RG running), which is obvious for $G$ but restrictive for $\Lambda$. Another also mild requirement was the additional kinetic terms to vanish when $G$ and $\Lambda$ turn to constants and thus to recover the Einstein equations. Finally, it was demanded that the coefficients of the covariant derivatives of $G$, $T_{\mu\nu}$ and $\Lambda$ to be determined from the Bianchi identities. This rules out the possibility of extra terms that contain $T$ or $T_{\mu\nu}$. One can also prove that, for consistency, the extra kinetic terms should have an even number of covariant derivatives for $G$ and $\Lambda$ while for avoiding known complications associated with higher-order derivatives, up to second-order derivatives need to appear in the modified equations. In the end, it has been proven that only kinetic terms of the cosmological constant survive, while all the possible kinetic terms of Newton's constant vanish. Thus, the modified Einstein equations read\footnote{For the derivation, check \cite{Bonanno:2020qfu}.}
\begin{align}\label{mod-Einstein-equ}
    G_{\mu\nu} = 8\pi G T_{\mu\nu}- \bar{\Lambda}e^{\psi}g_{\mu\nu} - \frac{1}{2}\nabla _{\mu}\psi \nabla _\nu \psi - \frac{1}{4} g_{\mu\nu} (\nabla \psi)^2+ \nabla _\mu \nabla _\nu \psi - g_{\mu\nu} \square \psi\,,
\end{align}
while the conservation equation is
\begin{align}\label{mod-conservation-equ}
    \nabla ^\mu (G T_{\mu\nu}) + G \left(T_{\mu\nu} - \frac{1}{2}Tg_{\mu\nu} \right) \nabla ^\mu \psi = 0\,.
\end{align}
In the above, $G_{\mu\nu}$, is the Einstein tensor, i.e. $G_{\mu\nu} = R_{\mu\nu} - \frac{1}{2}R g_{\mu\nu}$, $T_{\mu\nu}$ is the energy momentum tensor of the matter fields and $\psi$ is a \textit{determined} by RG flow scalar function (not a dynamical field), associated with the cosmological constant, i.e. $\Lambda = \bar{\Lambda} e^{\psi}$, where $\bar{\Lambda}$ is an arbitrary reference value with dimensions of inverse length squared. Note also, that the conservation equation \eqref{mod-conservation-equ} allows the interaction and exchange of energy among $T_{\mu\nu}$, $G$ and $\Lambda$ in this context.

It is more convenient to parameterize $G$ as well using the dimensionless parameters, in the same manner we did with $\Lambda$ 
\begin{equation}
    G = \bar{G} e^{\chi }\,,
\end{equation}
where $\bar{G}$ (with dimensions length squared) is arbitrary constant reference value. $T_{\mu\nu}$ obeys its own equations \eqref{mod-conservation-equ}, while $G$ and $\Lambda$ on the other hand, or better $\chi$ and $\psi$ respectively, are considered as predetermined spacetime functions, with a known behaviour, determined from the RG flow, and thus they do not obey some equations of motion. 

Before we proceed it is interesting to notice that, the present study differs significantly from the work of Reuter and Weyer \cite{Reuter:2003ca}; the consistency equation here, meaning the equation arising from satisfying the Bianchi identities, is very different. The modified Einstein equations have been determined allowing all possible up to second order covariant derivatives of $G$ and $\Lambda$ and then, we simply demand all coefficients to be determined in an \textbf{independent} way  from a particular choice of the background geometry or a specific choice of matter or specific functions of $G(k)$ and $\Lambda(k)$. Remarkably, all coefficients are uniquely determined, see \cite{Bonanno:2020qfu}, and there is no need for ad-hoc selections of the allowed set of covariant derivatives of $G$ and $\Lambda$ or of $T_{\mu\nu}$. Another difference is that the action we present here, couples $G(x^{\mu})$ to the matter.

\section{Constructing the action from the Equations of Motion}
\label{sec:constructing-action}

Let us know try to construct a suitable action, which after variation with respect to the metric will yield the above equations of motion \eqref{mod-Einstein-equ}. Again, we stress that, variations with respect to the scalar functions $\psi$ and $\chi$ are not considered, since these functions are not dynamical fields by construction i.e. they behave like coupling constants. Obviously, since the Einstein tensor is present, and specifically it is not coupled to another field, the Ricci scalar should appear in the action. However, the last two terms of Eq.~\eqref{sec:intro}, i.e.
\begin{equation}
    \nabla _\mu \nabla _{\nu} \psi \quad \text{and} \quad g_{\mu\nu}\square \psi \,,
\end{equation}
arise in scalar tensor theories from a non-minimal coupling of the scalar field to gravity. The fact that we need a non-minimal coupling of $\psi$ to the Ricci tensor, but still the Einstein tensor should appear linearly in the equations of motion, means that in the Lagrangian we should have an overall coupling of the dilatonic form
\begin{equation}
    \mathcal{L} \sim  f(\psi) \left[ R + ...\right]\,.
\end{equation}
The third and fourth term in Eq.~\eqref{mod-Einstein-equ}, i.e.
\begin{equation}
    - \frac{1}{2}\nabla _{\mu}\psi \nabla _\nu \psi \quad \text{and} \quad - \frac{1}{4} g_{\mu\nu} (\nabla \psi)^2\,,
\end{equation}
are easily obtained from a simple kinetic term of $\psi$ in the Lagrangian, that is of the form 
\begin{equation}
    \mathcal{L} \sim \omega (\nabla \psi)^2\,.
\end{equation}
Finally, the first term $8\pi G T_{\mu\nu}$ comes from a non-minimal coupling of the matter Lagrangian with Newton's constant, while the second one $\bar{\Lambda} e^\psi g_{\mu\nu}$ is of the potential form, $V(\psi)$. Thus, we end up with the action of the form
\begin{equation}\label{Lagrangian}
    \mathcal{S} =  \int \dd ^4 x\sqrt{-g}   f(\psi) \left[ R - \omega (\nabla \psi)^2 - V(\psi)  + 16 \pi G(x) \mathcal{L}_{\rm matter} \right]  \,.
\end{equation}
If we want to reproduce the exact form of Eq.~\eqref{mod-Einstein-equ} we have to set $f(\psi ) = e^\psi, \omega = - \frac{3}{2} $ and $ V(\psi) = 2 \bar{\Lambda}e^\psi $\,. 


It is interesting to notice that, the matter action couples to a running Newton's ``constant''. This is similar to the non-minimal matter-gravity coupling suggested by Markov and Mukhanov \cite{Markov:1985py,Zholdasbek:2024pxi}, but instead of the coupling to depend on the energy scale, here it depends on the spacetime coordinates. Variations of the Lagrangian \eqref{Lagrangian} with respect to $g^{\mu\nu}$ will give 
\begin{align}\label{eom}
    G_{\mu\nu} = 8\pi G T_{\mu\nu} - \frac{1}{2}V(\psi)g_{\mu\nu} + \left( \omega+ \frac{f''    }{f}\right) \nabla _{\mu}\psi \nabla _\nu \psi -  g_{\mu\nu}\left(\frac{\omega}{2} + \frac{f''}{f}\right) (\nabla \psi)^2 + \frac{f'}{f} \left(\nabla _\mu \nabla _\nu \psi - g_{\mu\nu} \square \psi\right)\,,
\end{align}
where 
\begin{equation}
     \frac{\delta \left(\sqrt{-g}16 \pi f(\psi)G(x)\mathcal{L}_{\rm matter} \right)}{\delta g^{\mu\nu}} = - 8 \pi \sqrt{-g} f(\psi)G(x)T_{\mu\nu}\,.
\end{equation}
As expected and in contrast with \cite{Reuter:2003ca}, variations of the action with respect to the metric do not contribute any derivative of $G(x)$ terms (like $\nabla _\mu \nabla _\nu G, \,\square G,$ etc) since the latter couples non-minimally only to matter fields. 

\section{Applications at different energy scales}\label{App}

In the present study, $G$ and $\Lambda$ are not considered independent dynamical fields, unlike Brans-Dicke like models and thus, they lack their own equations of motion. This is in agreement with RG approaches to Quantum Gravity, such as Asymptotic Safety, where $G$ and $\Lambda$ are treated as running coupling constants within an effective field theory approach. In this section, we will study the above action in different scenarios, namely in Asymptotic Safety for quantum cosmology, as well as for classical cosmology, where we consider two different scaling laws for the running of $G$ and $\Lambda$.

\subsection{Implementing  scaling law around fixed points}
Let us consider explicit models  $G$ and $\Lambda$ running we rewrite the action \eqref{Lagrangian} as
\begin{equation}\label{LagrangianDZ}
    \mathcal{S} = (\bar{\Lambda})^{-1}  \int \dd ^4 x \sqrt{-g}\,\Lambda\, \left[  R +\frac{3\,\Lambda^{-2}}{2} (\nabla \Lambda)^2 - 2 \,\Lambda  + 16 \pi G(x) \mathcal{L}_{\rm matter} \right] \,,
\end{equation}
where $\Lambda = \bar{\Lambda}\,e^{\psi(x)}$. This is an Einstein Hilbert like action with an extra kinetic term for the varying $\Lambda$. We immediately observe that in front of the kinetic term in action Eq.~\eqref{LagrangianDZ} there is a positive sign which would generate a problem if $\psi$ was a dynamical field. However, in our case as we have explained $\psi$ is not a dynamical field so there are no loops with respect to it.

There are two possibilities regarding this action. In the first approach we could study action Eq.~\eqref{LagrangianDZ} as a fundamental modified gravity action, like in the previous subsection. In the second more favorable approach we view action Eq.~\eqref{LagrangianDZ}, as an effective  action that could arise from Einstein-Hilbert action when vacuum polarization effect at the level of $G$ and $\Lambda$ are included. 
Vacuum polarization effect can be obtained withing AS and in this case we can assume we extract this information from the beta functions that provide the evolution of $G(k)$ and $\Lambda(k)$ with respect to an energy scale $k$. Let's suppose that the solution of the RG system of differential equations is $G=G(k)$ and $\Lambda=\Lambda (k)$, \cite{Reuter:2004nx}. Then, it is fair to work with action Eq.~\eqref{LagrangianDZ} using these solutions for $G(k),\,\,\Lambda(k)$. 

We can distinguish three interesting energy scales. First in the far UV limit, in the vicinity of the Reuter fixing point or otherwise called non Gaussian Fixing Point (NGFP), $G$ and $\Lambda$ are given by 
\begin{equation}
 G(k)=k^{-2}\,g_*\,\,\,\ \text{and}\,\,\, \Lambda=k^2\,\lambda_*,    
\end{equation} 
with $g_*$, and $\lambda_*$ being dimensionless constants. For lower energy scale we have
\begin{equation}
 G(k)=k^{-2}\,g(k)\,\,\,\ \text{and}\,\,\, \Lambda=k^2\,\lambda (k),    
\end{equation} 
where the dimensionless strength of curving and stretching the spacetime  $g(k)$, $\lambda(k)$ respectively, are solutions of the beta functions' partial differential equations. These beta functions can be calculated as in the seminal paper of Reuter \cite{Reuter:1996cp} and in \cite{Dou:1997fg} using a heat kernel type of expansion \cite{Zarikas:1999bf} for the functional traces or using a spectral sum over a spherical background \cite{Benedetti:2011ct}.

The scaling of $g(k)$ and $\lambda(k)$ reads \cite{Adeifeoba:2018ydh},
\begin{subequations}\label{flow}
	\begin{align}
		&g(k)=g_*+g_1\left(\frac{k}{M_{d}}\right)^{-\theta_1}+g_2\left(\frac{k}{M_{d}}\right)^{-\theta_2}\,,\\
		&\lambda(k)=\lambda_*+\lambda_1\left(\frac{k}{M_{d}}\right)^{-\theta_1}+\lambda_2\left(\frac{k}{M_{d}}\right)^{-\theta_2}\,,
	\end{align}
\end{subequations}
It is obvious that the critical exponents $\theta_i$ are important for the flow behaviour. A new energy scale is introduced unavoidably. 
At some energy scale $M_d$ the ensemble of quantum spacetimes decouple and provide the emergence of a classical spacetime. This $M_d$ is the maximum energy scale where we can trust the system of equations \eqref{mod-Einstein-equ}, \eqref{mod-conservation-equ} and \eqref{flow}. $M_d$ is expected to be related with the Planck energy scale $M_P$. 

Finally, there is another interesting energy scale close to the so called Gaussian Fixing Point, (GFP).
In this regime the RG flow can be linearized about
the GFP,
\cite{Bonanno:2007wg}, and we get
\begin{equation} \label{gfp}
 G(k)\simeq G_N\,\,\,\ \text{and}\,\,\, \Lambda(k) \propto k^4.    
\end{equation} 
This behaviour of $G$ and $\Lambda$ arises also using conventional quantum perturbation
theory. It has been shown in the context of AS theory, that there are trajectories that start from the NGFP, which is an attractive fixed point, and crossover the GFP towards the IR regime.

\subsection{Quantum Cosmology}
In this section, we substitute $\phi = e^\psi$ in action, (\ref{LagrangianDZ}), in order the notation to be similar with existing descriptions of a quantum cosmology scenario. We follow the Bryce de Witt reasoning , trying to discover possible new features, for example, how the running of $\Lambda$ can be consistent in quantum cosmology or what new terms appear in the equation in the mini-supersapce approximation.

The action, (\ref{LagrangianDZ}), takes the form
\begin{equation}\label{BDforquantum}
    \mathcal{S} = \int \dd ^4 x\, \sqrt{-g} \left[\phi R + \frac{3}{2\phi} (\nabla \phi)^2 - 2\Bar{\Lambda}\,\phi^2 + 16 \pi G(x) \phi \mathcal{L}_{\rm matter} \right] \,.
\end{equation}
The motivation to derive this canonical quantisation assuming a wave function of the whole Universe in a mini-superspace approach is solely to explore and identify differences with the same type of quantisation applied to Brans-Dicke (BD) action, \cite{Pal:2016hxt},\cite{Pimentel:2000ib}, that differs by the term $-2\Bar{\Lambda}\,\phi^2$ and the matter term. Note that in our case $\phi$ is a re-parameterisation of the varying $\Lambda$ and is not an independent field related to $G$ with its own equations of motion like in BD approach.
In order to proceed further let's work with the assumption that we are in a regime where $G$ varies very slowly. This could be expected to be valid below Planck scale in an asymptotic safety cosmological scenario. Thus, we refer to an energy scale described by equations (\ref{flow}) with critical exponents such to ensure a negligible variation of $G$ or to an even lower energy scale near the GFP with equations (\ref{gfp}).

Then, the previous action (\ref{BDforquantum}), can be written as  
\begin{equation}\label{BDforquantum2}
    \mathcal{S} = \int \dd ^4 x\,\sqrt{-g}  \left[\phi (R- 2\Bar{\Lambda}\,\phi ) + \frac{3}{2\phi} (\nabla \phi)^2 + 16 \pi G_N \phi \mathcal{L}_{\rm matter}\right] \,.
\end{equation}
We work with a homogeneous and isotropic metric with a spatially flat topology given by
\begin{equation}\label{metric}
ds^{2}=-N(t)^{2}dt^{2} + a^{2}(t)\left[dx^{2}+dy^{2}+dz^{2}\right]\,,
\end{equation}
where $N(t)$ is the lapse function and $a(t)$ is the scale factor. 
We include also in the matter sector a simple perfect fluid with equation of state, $p=W\rho$. So, the Lagrangian under study becomes
\begin{equation}\label{lag}
L= \phi (R- 2\Bar{\Lambda}\,\phi )+ \frac{3}{2\phi}\partial_{\mu}\phi\partial^{\mu}\phi + w\,\phi\rho,
\end{equation}
where $\phi$ is manifestly non-minimally coupled with the Ricci scalar and $w=16 \pi G_N\, W =\text{constant}$. In this context of an isotropic homogeneous universe, it is natural to work with a $\Lambda$ being a function of time, $t$, only. 
The Ricci scalar becomes 
\begin{equation}
    R=  \frac{6}{N(t)^2} \left[ \frac{\ddot{a}(t)}{a(t)} + \left( \frac{\dot{a}(t)}{a(t)} \right)^2 \right]
\end{equation}

We parameterize the scale factor and the scalar $\phi$ as $a(t) = e^{\kappa(t)}$, $\phi(t)= e^{\psi(t)}$. Then, 
$R=\frac{6}{N(t)^2} \left(2 \kappa'(t)^2+\kappa''(t)\right)$. At this point we have to make for simplicity the gauge fixing choice $N(t)=1$. This introduces well-known subtleties in the context of minusupersace quantum cosmology. The extra term $-2\Bar{\Lambda}\,\phi^2$ introduces complexity and the lapse function cannot be a common factor of the whole gravitational Lagrangian. Thus, the gauge fixing is needed at this point. 

Then, the gravitational part of the Lagrangian \eqref{lag} becomes
\begin{equation}
L_{\rm grav}=\frac{1}{2} e^{3 \kappa(t)+\psi (t)} \left(24 \kappa'(t)^2+12 \kappa''(t)-4 \bar{\Lambda}  e^{\psi (t)}-3 \psi '(t)^2\right)
\end{equation}
By introducing a new field $b(t)$ to replace 
$\kappa(t)$, using $\kappa(t)=b(t) -\psi(t)/2$, the above Lagrangian becomes  
\begin{align}
L_{\rm grav} &=  \frac{1}{2} e^{3 b(t)-\frac{\psi (t)}{2}} \left[3 \left(-8 b'(t) \psi '(t)+8 b'(t)^2+4 b''(t)+\psi '(t)^2
-2 \psi ''(t)\right)-4 \bar{\Lambda}  e^{\psi (t)}\right] \nonumber \\
&=   e^{3 b(t)-\frac{\psi(t)}{2}} \left( -6 b'(t)^2  -2 \bar{\Lambda}  e^{\psi (t)} \right) + {\rm total\, derivatives}.\label{Lgrav}
\end{align}

We can now observe differences with other quantisation schemes of Brans-Dicke actions. First of all, there is no kinetic term attributed to $\psi$ because we work in the case of $\omega=-3/2$. This results to the vanishing of $\psi'(t)^2$ terms. Second the extra $-2\Bar{\Lambda}\,\phi^2$ term in the action Eq.~\eqref{BDforquantum2}, generates the second term in the Lagrangian Eq.~\eqref{Lgrav}. The most important outcome of this result is that there is no canonical momentum associated with the $\psi(t)$ or equivalently with $\Lambda(t)$. This is consistent with the interpretation that $\Lambda$ is not a field with its own equations of motions like in BD theory. If instead we were working with the action of Eq.(\ref{Lagrangian}) with $\omega \neq - 3/2$ then a term $\psi'(t)^2$ would remain in the gravity Lagrangian resulting to a canonical momentum related to $\psi$ which would not be compatible with a coupling like interpretation of $\Lambda$. 
Therefore, now it is allowed to have a $\Lambda$ function of $k$ and in cosmological context a function of time, determined by AS theory, see \cite{Bonanno:2020qfu}. This indicates the significance and consistency of modified Einstein equations given in Eq.~\eqref{mod-Einstein-equ} as an effective AS gravity model.

The canonical conjugate momentum from the gravity sector is
\begin{equation}
    \pi_b=\frac{\partial L_{\rm grav}}{\partial b'}=-12b'\, e^{3b-\psi/2}\,,
\end{equation}
and the corresponding Hamiltonian is
\begin{equation}\label{Hg}
H_{\rm grav} = -\,\frac{1}{24} e^{\frac{\psi}{2}-3b}\,\pi_b^2 +2 \Bar{\Lambda}  e^{3b+\psi/2}.
\end{equation}
The Wheeler-DeWitt (WDW) equation arises with a canonical quantisation of the Hamiltonian constraint $H\Psi=0$. The operator in position space that we use is $\pi_b\rightarrow -i\frac{\partial} {\partial b}$. Finally,  we get using the Hawking-Page suggestion, \cite{Hawking:1985bk}, for ordering 
\begin{equation} \label{WDW1}
    \left[ e^{-b}\frac{\partial} {\partial b}(e^{-b}\frac{\partial} {\partial b}) +48\Bar{\Lambda}e^{4b} \right]\Psi=0
\end{equation}
We observe that $\psi$ does not enter the WDW equation but only through $b$.

If we want to include matter, we add a perfect fluid. In this case the Hamiltonian for the matter sector \cite{Pal:2016hxt}  is 
\begin{equation}
\label{3.93}
H_{\rm matter}= e^{3\psi(\frac{\psi}{2}-b)\,w}p_{T},
\end{equation}
where $p_{T}$ is the momentum associated with fluid. It has been suggested in \cite{Pal:2014dya} and \cite{Pal:2015vta}, to use the fluid term to define a time variable $T$ and conjugate momentum $p_{T}$. Then, we define a new operator with $p_{T}\mapsto-\imath\partial_{T}$,
Now, the WDW equation becomes
\begin{equation} \label{WDW2}
    \left[ e^{-b}\frac{\partial} {\partial b}(e^{-b}\frac{\partial} {\partial b}) +48\Bar{\Lambda}e^{4b} \right]\Psi=24\,e^{3w\psi}\,i\,\frac{\partial} {\partial T} \Psi
\end{equation}
A new feature in our case compared with similar studies is the appearance of this extra $\phi$ that multiplies the momentum $p_T$.
To simplify further the equation we define $z=e^b$, then
\begin{equation} \label{WDW3}
    \left[ \frac{\partial^2} {\partial z^2} +48\Bar{\Lambda}z^{4} \right]\Psi=24\,e^{3w\psi}\,i\,\frac{\partial} {\partial T} \Psi
\end{equation}
Using the ansatz that the solution is of the form $\Psi=f(z)\,e^{i\,E\,T}$, we have just to solve the differential equation
\begin{equation}
    f''(z)+48\bar{\Lambda} z^4 f(z)+24\,e^{3\psi\,w}\, E\,f(z)=0
\end{equation}
which can be solved numerically. For example, numerical solutions can be found working with particular RG flow for $\psi$. It is beyond the scope of this work to further analyze the space of solutions.

\subsection{Cosmology}
In this subsection we present some interesting cosmological solutions, some of which were presented also in \cite{Bonanno:2020qfu}, and are relevant for the perturbation analysis that follows in Section \ref{PPS}. The cosmology derived from the new Einstein and conservation equations (\ref{mod-Einstein-equ}), (\ref{mod-conservation-equ}) can be obtained assuming a spatially homogeneous and isotropic metric of the form
\begin{equation}
ds^{2}=-dt^{2}+a(t)^{2}\Big[\frac{dr^{2}}{1\!-\!\kappa\,r^{2}}+r^{2}\big(d\theta^{2}
\!+\!\sin^{2}{\!\theta}\,d\phi^{2}\big)\Big]\,, \label{cosm}
\end{equation}
with $\kappa=-1,0,1$ parameterising the spatial curvature options. 
Furthermore, we define the energy-momentum tensor $T^{\mu}_{\nu}$, to represent a fluid with energy density
$\rho$ and pressure $P$
\begin{equation}
T^{\mu\nu}=(\rho+P) u^{\mu}u^{\nu}+P g^{\mu\nu}\,,
\label{tmn}
\end{equation}
with $u^{\mu}$ the fluid 4-velocity and.

Finally the modified Friedman, Raychaudhuri and conservation equations are
\begin{gather}
H^{2}+\frac{\kappa}{a^{2}}\!=\!\frac{\bar{\Lambda}}{3}e^{\psi}-H\dot{\psi}
-\frac{\dot{\psi}^{2}}{4}+\frac{8\pi}{3}G\rho\,,
\label{cosmo1} \\
\dot{H}=\frac{\kappa}{a^{2}}+H\frac{\dot{\psi}}{2}+\frac{\dot{\psi}^{2}}{4} -\frac{1}{2}\,\ddot{\psi}-4\pi G (\rho+P)\,,
\label{cosmo2} \\
\dot{\rho}+3H(\rho+P)+\rho\,\dot{\chi}+\frac{\rho+3P}{2}\dot{\psi}=0\,,
\label{cosmo3}
\end{gather}
where $H=\dot{a}/a$ is the Hubble parameter (dot denotes derivative with respect to $t$). Assuming a constant equation of state parameter for simplicity, $w=\frac{P}{\rho}$, the conservation equation \eqref{cosmo3} and the demand equation \eqref{cosmo2} to be redundant with equation \eqref{cosmo1} gives
\begin{equation}
G\rho=\frac{c}{a^{3(1+w)}}\,e^{-\frac{1+3w}{2}\psi}\,.
\label{cosmo4}
\end{equation}

In this subsection we will study the early cosmic evolution of some interesting cosmological solutions of equations \eqref{cosmo1}, \eqref{cosmo3}. From now one we work with a flat Universe $\kappa = 0$. To proceed further we will assume that the  cosmology era is described from an AS gravity model near the Non Gaussian Fixed point of the RG flow evolution. Since $\Lambda=\bar{\Lambda}\,e^{\psi}$ and $G=\bar{G}\,e^{\chi}$ one can recast $\psi$ and $\chi$ as 
\begin{equation}
\psi = \ln\left(\frac{\lambda_* k^2}{\bar{\Lambda}}\right),\quad \chi = \ln\left(\frac{g_*}{\bar{G}\,k^2}\right).
\end{equation}
In the context of AS phenomenology it is required at this point to connect the $k$ scale of running with a physical scale. The geometry-independent RG flow equations include running $\Lambda$ and $G$ in terms of a characteristic scale $k$, which can be associated to a characteristic time or length scale of the system at hand. In particular, for the cosmological setup, it is understood that $k$ should be related with the cosmic time. Since AS theory cannot provide this relation researchers usually make use of arbitrary functions of of $t$, $a$, $H$, $\rho$, curvature etc. In the present study we will use two popular scaling laws $k\propto 1/t$ and $k\propto H$.

\subsubsection{scaling \texorpdfstring{$k\propto 1/t$}{kt}}
One popular choice for the cutoff $k$ is to set 
\begin{equation}
k=\frac{\xi}{t}    
\end{equation}
where $\xi$ a dimensionless contant expected to be of order of one. Since in the very early Universe the physics is not well known we will distinguish various cases. According to AS theory, the value of $\Lambda$ goes to infinity for transplanckian energies. One case is to set the energy momentum tensor equal to zero. Afterwards we will also present a solution without disregarding $T^{\mu}_{\nu}$. In this scaling, $e^{\psi}=\lambda_{\ast}\xi^{2}/(\bar{\Lambda}t^{2})$ and thus equation  \eqref{cosmo1} becomes
\begin{equation}
H^{2}=\Big(\frac{\lambda_{\ast}\xi^{2}}{3}\!-\!1\Big)\frac{1}{t^{2}}
+\frac{2H}{t}\,.
\label{drlvy}
\end{equation}
the solution is
\begin{equation}
a(t)=C\, t^{1 \pm \frac{\sqrt{\lambda_{*} }\, \xi }{\sqrt{3}}}
\end{equation}
where $C$ an integration constant. We observe that one branch of the solutions give power law inflation i.e. $a(t)\propto t^p$ with $p>1$.

Another case is to include the contribution of the radiation. In this case $w=1/3$ and the Friedman equation \eqref{cosmo1} becomes
\begin{equation}
H^{2}=\Big(\frac{\lambda_{\ast}\xi^{2}}{3}\!-\!1\Big)\frac{1}{t^{2}}
+\frac{2H}{t}+\frac{8\pi c\bar{\Lambda}t^{2}}{3\lambda_{\ast}\xi^{2}a^{4}}\,.
\label{drlvy2}
\end{equation}
 Following \cite{Bonanno:2020qfu} we define
\begin{equation}
\varpi=\sqrt{\frac{\lambda_{\ast}\xi^{2}}{3}}\,\,\,\,\,,\,\,\,\,\,
\sigma=\sqrt{\frac{8\pi c\bar{\Lambda}}{3\lambda_{\ast}\xi^{2}}}\,\,.
\label{gywg}
\end{equation}
where the constant
$\sigma$ is a reparametrization of the integration
constant $c$. 
Then, the solution is 
\begin{equation}
a(t)=\frac{\sqrt{\sigma}\,\,t}{\sqrt{2\varpi}}\,\sqrt{\Big(\frac{t}{t_{o}}\Big)^{\!\pm 2\varpi}
\!-\!\Big(\frac{t_{o}}{t}\Big)^{\!\pm 2\varpi}}\,,
\end{equation}
where the upper $\pm$ branch holds for $t>t_{o}$, while the lower one for $t<t_{o}$, with $t_{o}$ a positive integration constant. The interesting solution is the upper branch which gives an accelerating solution starting with $a=0$ at $t=t_0$. The scale factor approximately behaves after some time like 
\begin{equation}
a(t)=\frac{\sqrt{\sigma}\,\,}{\sqrt{2\varpi}}\,\Big(\frac{1}{t_{o}}\Big)^{\!+ \varpi}\,\,t^{1+\varpi}
\,,
\end{equation}
This is also a power law inflation type of expansion.

The previous solutions are relevant for energies that are described with a running of $G$ and $\Lambda$ near the Non Gaussian fixing point (NGFP). For lower energies, in the GFP regime, \cite{Bonanno:2010mk, Mandal:2022aeq}, the Newton constant $G(k)$ is almost constant while $\Lambda$ run as $\Lambda(k) \propto k^4$ till also a constant value. 
In this regime, the cosmic evolution in the case of insignificant contribution from the perfect fluid and setting $\Lambda(k) =\lambda\, k^4$ is governed by the following expansion equation:
\begin{equation}
H^2=\frac{4 H}{t}+\frac{\lambda  \xi ^4 }{3 t^4}-\frac{4 }{t^2}
\end{equation}
with solution
\begin{equation}
   a(t)=c \, t^2 e^{\pm\frac{\sqrt{\lambda } \xi ^2}{\sqrt{3} t}}
\end{equation}
where $c$ an integration constant. It is easy to check that both branches provide inflationary expansion. The upper branch initially contracts rapidly while later expands like $a\propto t^2$, i.e. power law inflation. The lower branch is always expanding and after some time gives power law inflation.

Let us now include a perfect fluid in this study of the regime with constant $G(k)$ and $\Lambda(k) =\lambda\, k^4$.
with equation of state $P=w\,\rho$. Then using equations (\ref{cosmo1},\ref{cosmo4}), we get as expansion rate the following expression 
\begin{equation}
    H^2 = \frac{4 H}{t}+\frac{8}{3} \pi\,  c\, a^{2-3 (w+1)} \left(\frac{\lambda\,  \xi ^4}{\bar{\Lambda}\, t^4}\right)^{-\frac{1}{2} (3 w+1)}+\frac{\lambda\,  \xi^4 a^2}{3\, t^4}-\frac{4 a^2}{t^2}\,.
\end{equation}
However, this differential equation can be solved only numerically even if we set $w=1/3$ or $w=0$.  

\subsubsection{scaling  \texorpdfstring{$k\propto H$}{kH}}

An alternative and favorite cutoff choice can be formulated using the Hubble scale, 
$H(t)$  \cite{Reuter:2005kb}, 
\begin{equation}
k=\xi H(t)\,,
\label{scale2}
\end{equation}
where $\xi$ a positive constant. In our case since we have the inverse of $k$ to represent a characteristic length scale over which the averaging procedure is estimated, the positivity of $k$ and therefore of $H$ is modeled. In cosmology  $H$ can change sign but since in our case we describe the early Universe expansion we can make this reasonable modeling using equation \eqref{scale2}.
From this equation and working in the NGFP, we get $e^{\psi}=\lambda_{\ast}\xi^{2}H^{2}/\bar{\Lambda}$. 
The vacuum equation for expansion, for $\kappa=0$ is 
\begin{equation}
(1\!-\!\varpi^{2})H^{2}+2\dot{H}+\frac{\dot{H}^{2}}{H^{2}}=0\,,
\label{vac}
\end{equation}
The solution is 
\begin{equation}
  H(t)=\frac{1}{t (1\pm\varpi )+C}
\end{equation}
Thus, the scale factor is given by
\begin{equation}\label{non-singular-scale-factor-without-matter}
  a(t)= \Tilde{C} \left[C+ t(1+\delta\, \varpi) \right]^{\frac{1}{1+\delta \,\varpi }}  \,.
\end{equation}
with $\delta=\pm 1$ and $C$ and $\Tilde{C}$, integration constants.
Here we can have a non singular solution with power low inflation.

If we include matter in the form of radiation $w=1/3$
then, the Friedmann equation for flat topology is
\begin{equation}
(1\!-\!\varpi^{2})H^{2}+2\dot{H}+\frac{\dot{H}^{2}}{H^{2}}-\frac{\sigma^{2}}{a^{4}H^{2}}=0\,,
\label{hyd}
\end{equation}
The continuity equation (\ref{cosmo4}) for $\rho$ takes the form
\begin{equation}
\rho=\frac{9\varpi^{2}\sigma^{2}}{8\pi g_{\ast}\lambda_{\ast}a^{4}}\,.
\label{H2}
\end{equation}
Now it not possible to solve analytically for $a(t)$ in the most general case. Instead we can rewrite equation (/\ref{H2}), as
\begin{equation}\label{H22}
    \frac{d(\dot{a}^2)}{da}=\pm \frac{2}{a} \sqrt{\varpi^2\,\dot{a}^4+\sigma^2 }
\end{equation}
Since we are interesting to describe expanding and accelerating early universe cosmology we disregard branches of solutions with decreasing scale factor. 
Then we integrate equation \ for $\dot{a}^2$. The integral gives an inverse hyperbolic tangent function of $\dot{a}^2$ and using the identity 
  $1-\frac{2}{x^2+1}=\tanh[\log(x)]$ we can solve for the Hubble rate. Finally,we get 
\begin{equation}
    H^2=\frac{\sigma}{2\varpi}\Big |  (\frac{a}{a_c})^{2\varpi+2}-(\frac{a_c}{a})^{2\varpi-2}  \Big |
\end{equation}
where $a_c$ a positive integration constant.

For value of $a$ close to zero and $\varpi>1$ we have $H^2 \simeq \frac{\sigma}{2\varpi}(\frac{a_c}{a})^{2\varpi-2}$ so in this case we can find a simple expression for the scale factor as a function of time.
We find non singular expanding solutions:
\begin{equation}
    a(t)=\left[(\varpi -1 ) a_c^{\varpi } \left(\frac{\sqrt{\sigma } }{\sqrt{2\varpi}\, a_c}\,t+C\right)\right]{}^{\frac{1}{\varpi -1}}
\end{equation}
where $C$ integration constant. These non singular expanding solutions can also give power law inflation for $1<\varpi<2$. The same case with $a \rightarrow 0$ but for $\varpi<1$ gives initially decreasing scale factor.

For energies quite below Planck scale, we can assume that we are in the GFP, i.e. the Newton's constant is almost constant and $\Lambda \propto k^4$. In this case again like in the previous scale $k\propto1/t$ there are no analytical solutions and we have to numerically solve equation (\ref{cosmo1}).

In summary, we have analysed some interesting expanding cosmological flat solutions near the NGFP.
We have studied both vacuum, meaning $\Lambda$ dominated Universe and perfect fluid content 
For the scaling $k\propto 1/t$ there are branches of solutions that give an expanding power law inflationary cosmic period both for the vacuum case, as well as for a cosmic radiation content. For the $k\propto H$ scaling, we have discovered non singular power law inflationary expansions for both vacuum and radiation filled Universes.
Kretschmann invariant was evaluated for the non singular behaviour.

\section{Primordial Power Spectrum}
\label{PPS}

The action Eq.\eqref{Lagrangian} is not a sub-class of the Horndeski gravity. However, the vacuum action under consideration, looks exactly like a scalar-tensor theory in four dimensions, with a single scalar field that leads to second order field equations but only for the metric and not for the scalar field since the scalar function models a coupling and not a dynamical field. Here, we use the fact of this action similarity to utilize already known results of Horndeski gravity. However, one must be careful not to use any equations of motion for the scalar.
Specifically, the mapping of Eq. \eqref{Lagrangian} to Horndeski becomes with
\begin{equation}
    G_2(\psi,X) = f(\psi)(2\omega X - V(\psi))\,,\quad G_3(\psi,X) = 0\,,\quad G_4(\psi,X) = f(\psi)\,,\quad G_5(\psi,X) = 0\,,
\end{equation}
where $X = - (\nabla \psi)^2/2$ is the kinetic term of the scalar.

The quadratic action for tensor and scalar cosmological perturbations in Horndeski gravity is known. In particular, in the unitary gauge where the scalar field perturbations vanish, $\delta \psi = 0,$ and the metric is perturbed as
\begin{equation}
    \dd s^2 = - N^2 \dd t^2 + \gamma _{ij} (\dd x^i + N^i\dd t)(\dd x^j + N^j\dd t)\,,
\end{equation}
where 
\begin{equation}
    N = 1 + \alpha\,, \quad N_i = \partial _i \beta\,,\quad \gamma _{ij} = a^2(t) e^{2\zeta} \left( \delta _{ij} + h_{ij} + \frac{1}{2}h_{ik}h_{kj}\right)\,,
\end{equation}
we have the scalar perturbations $\alpha, \beta$ and $\zeta$ and the tensor perturbations $h_{ij}$ which is traceless and transverse, i.e. $h_{ii} = 0 = h_{ij,j}.$ In what follows, we consider only the gravitational perturbations and thus $T_{\mu\nu} = 0.$ Furthermore, the vector perturbations has been shown to be decaying in Horndeski gravity and thus are ignored here.

\subsection{Tensor perturbations}
The quadratic action of the tensor perturbations is 
\begin{equation}\label{eq:quadratic-tensor-action}
    \mathcal{S}_{\rm T} = \frac{1}{8}\int \dd t \dd ^3x a^3 \left[ \mathcal{G}_{\rm T} \dot{h}^2_{ij} - \frac{\mathcal{F}_{\rm T}}{a^2} (\vec{\nabla} h_{ij})^2\right]\,,    
\end{equation}
where 
\begin{align}\label{eq:tensor-coeff}
    \mathcal{F}_{\rm T} = 2 G_4  = 2 f(\psi)\,,\quad   \mathcal{G}_{\rm T} =  2 G_4 = 2 f(\psi)\,.
\end{align}
The squared sound speed of the perturbations is given by
\begin{equation}
    c_{\rm T}^2 = \frac{\mathcal{F}_{\rm T}}{\mathcal{G}_{\rm T}}\,,
\end{equation}
and thus it is always equal to unity. From \eqref{eq:quadratic-tensor-action} one can see that in order to avoid ghost and gradient instabilities we have to set respectively,
\begin{equation}
    \mathcal{G}_{\rm T} >0 \,,\quad \mathcal{F}_{\rm T}>0\,,
\end{equation}
which is satisfied if $f(\psi) > 0.$

We can perform the following change of variables
\begin{equation}
    \dd y_{\rm T} = \frac{1}{a}\dd t\,,\quad z_{\rm T} = \frac{a}{2}\sqrt{2f(\psi)}\,,\quad v_{ij} = z_{\rm T}h_{ij}\,,
\end{equation}
and we end up with the canonically normalized quadratic action for the tensor perturbations, that reads
\begin{equation}
    \mathcal{S} _{\rm T} = \frac{1}{2}\int \dd y_{\rm T }\dd^3x \left[(v'_{ij})^2 - (\vec{\nabla}v_{ij})^2 + \frac{z''_{\rm T}}{z_{\rm T}}v_{ij}^2\right]\,,
\end{equation}
where $'$ denotes differentiation with respect to $y_{\rm T}.$ The solution of the equations of motion of $v_{ij}$ at superhorizon scales are
\begin{equation}\label{eq:sol1}
    v_{ij} \propto z_{\rm T}\,,\quad v_{ij} \propto z_{\rm T}\int \frac{\dd y_{\rm T}}{z^2_{\rm T}}\,,
\end{equation}
which in terms of the non-canonical variables becomes
\begin{equation}\label{eq:sol2}
    h_{ij} = \text{const.}\,,\quad h_{ij} = \int ^t \frac{\dd t'}{2a^4 f(\psi)}\,.
\end{equation}

To evaluate the power spectral density we assume that
\begin{equation}
    \epsilon := - \frac{\dot{H}}{H^2} \simeq {\rm const.}\,,\quad f_{\rm T} := \frac{\dot{\mathcal{F}}_{\rm T}}{H\mathcal{F}_{\rm T} }= \frac{f'(\psi)}{f(\psi)} \frac{\dot{\psi}}{H} \simeq {\rm const.}\,,\quad g_{\rm T} := \frac{\dot{\mathcal{G}}_{\rm T}}{H\mathcal{G}_{\rm T} } = \frac{f'(\psi)}{f(\psi)} \frac{\dot{\psi}}{H} \simeq {\rm const.}\label{epsilon and g_T - tensors}
\end{equation}
We have to impose $\epsilon <1$ in order for $y_{\rm T}$ to run from $-\infty$ to $0$ as the Universe expands, while for the second solution of Eq.~\eqref{eq:sol1}-\eqref{eq:sol2} to decay we have to assume that $\epsilon - g_{\rm T} < 3.$ Solving the mode functions equation in Fourier space, we get
\begin{equation}
    v_{ij} = \frac{\sqrt{\pi}}{2} \sqrt{-y_{\rm T}}H_{\nu _{\rm T}}^{(1)} (-k y_{\rm T})\mathrm{e}_{ij}\,,
\end{equation}
with $H_{\nu _{\rm T}}^{(1)}$ being the Hankel function of first kind (plus sign), $\nu _{\rm T}$ being a positive scalar defined as
\begin{equation}
    \nu _{\rm T} := \frac{3 - \epsilon + g_{\rm T}}{2 - 2 \epsilon }\,,
\end{equation}
and $\mathrm{e}_{ij}$ the polarization tensor. Thus, the power spectrum of the primordial tensor fluctuations becomes
\begin{equation}
    \mathcal{P}_{\rm T} = 8 \gamma _{\rm T}  \frac{1}{\sqrt{2f(\psi)}} \frac{H^2}{ 4\pi ^2} \Big\rvert _{-k y_{\rm T} = 1}\,,
\end{equation}
where 
$$
\gamma _{\rm T} = 2^{2\nu_{\rm T} - 3}  \left|\frac{\Gamma (\nu _{T})}{\Gamma (3/2)}\right|^2 (1-\epsilon)\,.
$$
We evaluate the power spectrum at the sound horizon exit, i.e. $-ky_{\rm T} = 1$. The tensor spectral index is given by
\begin{equation}\label{tensor spectral index}
    n_{\rm T} = 3 - 2 \nu _{\rm T}\,,
\end{equation}
and the scale-invariant limit for tensor perturbations would be for $\nu _{\rm T} = 3/2$. From Eq.~\eqref{tensor spectral index} we see that, the gravitational wave spectrum could have a blue tilt if
\begin{equation}
    n_{\rm T} > 0 \Rightarrow 4 \epsilon + 2 g_{\rm T}  < 0\,.
\end{equation} 
 
We will apply the previous developed formalism  for the energy scale near the NGFP  and for the case of $k=\xi\,H$ scaling. We focus on the 
non singular vacuum solution derived for scaling 
The parameters in Eq.~\eqref{epsilon and g_T - tensors} for the scale factor given by \eqref{non-singular-scale-factor-without-matter} and the scalar $e^{\psi}=\lambda_{\ast}\xi^{2}H^{2}/\bar{\Lambda}$ become
\begin{equation}
 \epsilon = 1\pm \varpi \,,\quad f_{\rm T} = g_{\rm T} = -2 (1\pm \varpi)\,,   
\end{equation}
for $\delta =\pm 1$ and $\nu _{\rm T} = 3/2$ for both solutions and thus $n_{\rm T} = 0$. The fact that the tensor spectral index vanishes implies that the power spectrum of the tensor perturbations is scale-invariant, which in turn means that the amplitude of the primordial gravitational waves is the same across all scales or wavelengths. However, we have already assumed that $\epsilon < 1$ and $\epsilon - g_{\rm T} <3$ above and since $\varpi > 0$ from Eq.~\eqref{gywg}, only the solutions with $\delta = -1$ is in agreement with that.

\subsection{Scalar perturbations}
Similarly, the quadratic action for the scalar perturbations reads
\begin{align}\label{quadratic scalar action}
     S_{\rm S}^{(2)}=\int \dd t \dd^3 x a^3 \Bigg[\mathcal{G_S}\dot{\zeta}^2-\frac{\mathcal{F_S}}{a^2}(\vec{\nabla}\zeta)^2\Bigg] ,
\end{align}
where the coefficients read
\begin{gather}
\mathcal{G}_{\rm S} = \frac{\Sigma}{\Theta ^2}\mathcal{G}_{\rm T}^2 + 3 \mathcal{G}_{\rm T} \,, \\
\mathcal{F}_{\rm S} = \frac{1}{a}\frac{\dd}{\dd t}\left( \frac{a}{\Theta}\mathcal{G}_{\rm T}^2\right) - \mathcal{F}_{\rm T} \,.
\end{gather}
where 
\begin{gather}
    \Sigma = X G_{2X} - 6 H^2 G_4- 6 H \dot{\psi} G_{4\psi} = 2\omega Xf(\psi) - 6 H^2 f(\psi) - 6 H \dot{\psi} f'(\psi) \,, \\
    \Theta =  2 H G_4 + \dot{\psi} G_{4\psi} = 2 H f(\psi) + \dot{\psi}f'(\psi)\,,
\end{gather}
and $\mathcal{G}_{\rm T}, \mathcal{F}_{\rm T}$ are the coefficients of the tensor perturbations given in Eq.~\eqref{eq:tensor-coeff}.

In the same spirit as with the tensor perturbations, we define the new variables
\begin{equation}
    \dd y_{\rm S} = \frac{c_{\rm S}}{a}\dd t \,,\quad z_{\rm S} = \sqrt{2}a (\mathcal{F}_{\rm S}\mathcal{G}_{\rm S})^{1/4}\,,\quad u = z_{\rm S} \zeta\,,
\end{equation}
and the quadratic scalar action is canonically normalized to take the known ``Sasaki-Mukhanov'' form
\begin{equation}
    \mathcal{S}_{\rm S} = \frac{1}{2}\int \dd y_{\rm S} \dd x^3 \left[ (u')^2 - (\vec{\nabla}u)^2 + \frac{z''_{\rm S}}{z_{\rm S}}u^2\right]\,,
\end{equation}
where $'$ denotes differentiation with respect to the canonical time. 

The two independent solutions on superhorizon scales read
\begin{equation}
    \zeta = {\rm const.}\quad {\rm and} \quad \zeta \propto \int ^t \frac{\dd t'}{a^3 \mathcal{G}_{\rm S}}\,.
\end{equation}

Following the same procedure as with the tensor perturbations to evaluate the power spectrum in this case, we assume
\begin{equation}\label{epsilon, f_S and g_S}
   \epsilon := - \frac{\dot{H}}{H^2} \simeq {\rm const}\,,\,\, f_{\rm S}:= \frac{\dot{\mathcal{F}}_{\rm S}}{H \mathcal{F}_{\rm S}} \simeq {\rm const}\,,\,\, g_{\rm S} := \frac{\dot{\mathcal{G}}_{\rm S}}{H \mathcal{G}_{\rm S}} \simeq {\rm const}\,.
\end{equation}
Then, the power spectrum will be given by
\begin{equation}
    \mathcal{P}_{\rm S} = \frac{\gamma _{\rm S}}{2} \frac{\mathcal{G}_{\rm S}^{1/2}}{\mathcal{F}_{\rm S}^{3/2}} \frac{H^2}{4\pi ^2}\Big\rvert_{-ky_{\rm S} = 1}\,,
\end{equation}
where $$\nu _{\rm S}:= \frac{3 - \epsilon + g_{\rm S}}{2- 2\epsilon - f_{\rm S} + g_{\rm S}}$$ and $$\gamma _{\rm S} = 2 ^{2 \nu _{\rm S}-3} \left|\frac{\Gamma (\nu _{\rm S})}{\Gamma (3/2)}\right|^2(1-\epsilon - \frac{f_{\rm S}}{2}+\frac{g_{\rm S}}{2})\,.$$ 
The scalar spectral index is given by
\begin{equation}\label{scalar-spectral-index}
    n_{\rm S} = 4 - 2 \nu _{\rm S}\,,
\end{equation}
and thus a spectrum with equal amplitudes at horizon crossing should obey
\begin{equation}
    \epsilon + \frac{3 f_{\rm S}}{4} - \frac{g_{\rm S}}{4} = 0\,.
\end{equation}

If we consider the limit $\epsilon, f_{\rm T}, g_{\rm T}, f_{\rm S},g_{\rm S} \ll 1$, we get $\nu _{\rm T} , \nu _{\rm S} \rightarrow 3/2$ and thus $\gamma _{\rm T} , \gamma _{\rm S} \rightarrow 1$. The tensor-to-scalar ratio is then given by
\begin{equation}\label{eq:ratio}
    r = 16 \frac{\mathcal{F}_{\rm S}}{\mathcal{F}_{\rm T}} \frac{c_{\rm S}}{c _{\rm T}}\,.
\end{equation}

Up to leading order in slow-roll, we have from Eq.~\eqref{tensor spectral index}
$n_{\rm T} \simeq - (2\epsilon + g_{\rm T})$ with $g_{\rm T} = \dot{f}/(Hf),$ and $c_{\rm S}^2 = 1$ with
\begin{equation}
    \mathcal{F}_{\rm S}=\mathcal{G}_{\rm S} \simeq 2\omega f\frac{X}{H^2}  \simeq f(2\epsilon + g_{\rm T}) \simeq - f n_{\rm T}\,,
\end{equation}
which yields from Eq.~\eqref{eq:ratio} the standard consistency relation
\begin{equation}
    r \simeq - 8 n_{\rm T}\,.
\end{equation}

Similarly as before, we work for the energy scale near the NGFP  and for the case of $k=\xi\,H$ scaling. The scalar spectral index \eqref{scalar-spectral-index} for the non-singular scale factor \eqref{non-singular-scale-factor-without-matter} becomes
\begin{equation}
    n_{\rm S} = \frac{2}{1 \mp \varpi} \quad {\rm for} \quad \delta = \pm 1 \,.
\end{equation}
The parameters in Eq.~\eqref{epsilon, f_S and g_S} take now the form
\begin{equation}
    \epsilon = 1 \pm \varpi\,,\quad f_{\rm S} = -2 (1\pm \varpi) = - 2 g_{\rm S}\,,
\end{equation}
and since $\epsilon < 1$ and $\varpi >0$, only the $\delta = -1$ solution is in agreement with that. Remember that the scalar spectral index characterizes the scale dependence of the primordial density perturbations in the early universe. It provides information on how the amplitude of these perturbations varies with scale (or wavenumber) and its value according to Planck \cite{Planck:2018vyg} should be slightly less than 1, $n_{\rm S} = 0.965$ (red tilt). In order to achieve that in our current setup we have to consider $\varpi > 1$.

\section{Conclusions}

Asymptotic Safety framework suggests that gravity can be described as a fundamental quantum field theory. Proposed by Steven Weinberg in the 1970s, AS has garnered interest as a potential route to a consistent theory of quantum gravity. It has received great attention in recent times and we are going through an interesting productive period. More work is needed and is carried on regarding both the theory and its phenomenology, \cite{Anagnostopoulos:2018jdq, Anagnostopoulos:2022pxa, Kofinas:2017gfv}.

A few years ago, some of the authors of the current constructed a set of effective field equations that extend Einstein's equations and include a varying Newton's and cosmological constant. Motivated by AS and by the fact that the Bianchi identities should be satisfied, since the resulting theory should be diffeomorphism invariant, they derived Eq.~\eqref{mod-Einstein-equ} and \eqref{mod-conservation-equ} where $G$ and $\Lambda$ are functions of spacetime, which do not obey their own equations of motion, but instead, are determined by the RG flow.

Expanding on the above, in this work, we constructed a new gravitational action that generates these modified Einstein equations. The action consistently and uniquely encapsulates $\Lambda$ and $G$ as non propagating functions of spacetime and without problematic non local terms. This action was further RG improved in the context of Asymptotic Safety allowing $\Lambda$ and $G$ to be determined by an RG flow.  

Consequently, the RG improved action was applied in a quantum cosmology scenario  in order to check the new properties of this action and compare it with previous studies. The analysis reveal a consistent behaviour allowing the inclusion of a varying $\Lambda$ and at the same time without adding extra degrees of freedom and an extra conjugate momentum term. Furthermore, the action was tested on an isotropic and continuous spacetime giving an interesting non singular inflationary early cosmology.  

Last but not least, we have derived the most generic quadratic actions for tensor and scalar cosmological perturbations and used them to present stability criteria for both types of perturbations. Additionally, we have computed the primordial power spectra and applied them to the non singular cosmology that we have derived to constrain its free parameters. The allowed set of parameters describe a non singular cosmology with power law inflation. 

With increasingly precise measurements of CMB anisotropy, studying non-Gaussian signatures of primordial perturbations from inflation becomes crucial. To achieve this, we need as a future wowrk to compute the action up to cubic (and higher) orders in perturbations. It would also be interesting to explore the application of the proposed action in the context of scale dependent gravity \cite{Alvarez:2020xmk}.

\section*{Acknowledgements}
K.F.D. was supported by the PNRR-III-C9-2022–I9 call, with project number 760016/27.01.2023. This paper is based upon work from COST Action CA21136 {\it Addressing observational tensions in cosmology with systematics and fundamental physics} (CosmoVerse) supported by COST (European Cooperation in Science and Technology). We thank Prof. Pal Sridip for enlightening discussions. Author names in alphabetical order.

\bibliography{ref.bib}

\end{document}